\documentclass[epjCONF,columns]{svjour} 
\pdfoutput=1
\usepackage{graphics}
\usepackage[varg]{txfonts} 
\usepackage[latin1]{inputenc}
\usepackage{atlasphysics}

\session-title{Hadron Collider Physics Conference, Paris, 2011}

\def\zp{\ensuremath{Z'\;}}
\def\ifb{\mbox{fb$^{-1}$}}
\def\ipb{\mbox{pb$^{-1}$}}

\def\esix{\ensuremath{{E}_6\;}} 
\def\gstar{\ensuremath{G^*}}
\def\zppsi{\ensuremath{Z'_\psi}} 
\def\zpchi{\ensuremath{Z'_\chi}} 
\def\zpeta{\ensuremath{Z'_\eta}} 
\def\zpsq{\ensuremath{Z'_{\rm S}}} 
\def\zpN{\ensuremath{Z'_{\rm N}}} 
\def\zpI{\ensuremath{Z'_I}}

\def\xbr{\ensuremath{\sigma B\;}}
\def\antibar#1{\ensuremath{#1\bar{#1}}}
\def\ttbar{\antibar{t}}

\newcommand{\LimitCombinedGOne}{0.71}
\newcommand{\LimitCombinedGThree}{1.03}
\newcommand{\LimitCombinedGFive}{1.33}

\newcommand{\LimitCombinedG}{1.63}

\newcommand{\LimitCombinedSq}{1.60}
\newcommand{\LimitCombinedN}{1.52}
\newcommand{\LimitCombinedPsi}{1.49}
\newcommand{\LimitCombinedChi}{1.64}
\newcommand{\LimitCombinedEta}{1.54}
\newcommand{\LimitCombinedI}{1.56}

\begin{document}
%
\title{High Mass Resonances at ATLAS}
\author{Wojciech Fedorko\inst{1}\fnmsep\thanks{\email{wojtek.fedorko@cern.ch}} on behalf of the ATLAS Collaboration }
\institute{Michigan State University, Lansing, MI, USA}
\abstract{
A brief overview of searches for high mass resonances using a subset of data collected by the ATLAS experiment during the 2011 LHC run is presented. Various final states are explored including dilepton, diphoton, lepton with missing transverse energy, dijet, photon with a jet, top anti-top pairs, and $Z$ boson pairs. No new resonance has been found and limits on several new physics models are set. 
} 
\maketitle
\section{Introduction}
\label{sec:Introduction}

The Standard Model (SM) of particle physics has been stunningly successful in explaining experimental data. However several questions persist. At the time of writing the mechanism of electroweak symmetry breaking is unconfirmed, no viable candidate for dark matter particle has been observed and the hierarchy of masses of particles is unnatural within the SM. Several of the proposed solutions to these problems result in the production of heavy resonances with masses accessible at the LHC. In these proceedings an overview of several searches for such resonances performed at the ATLAS experiment~\cite{atlas:detector} is presented. Final states explored range from experimentally clean dilepton and diphoton states to quite challenging to analyze $t\bar{t}$ and diboson final states. Data presented here have been collected during the 2011 LHC run. 

\section{Search for a heavy dilepton resonance}

Heavy resonances decaying to a lepton-anti lepton pair are predicted by several extensions to the SM. These include neutral gauge boson \zp\cite{London:1986dk,Langacker:2008yv,Erler:2009jh}, techni-mesons~\cite{Lane:1989ej,Lane:2003,Belyaev:2009} and Kaluza-Klein (KK) excitations of the Graviton in the \\Randall-Sundrum (RS) model~\cite{RS}. In this analysis~\cite{US:Zprime2011} the Sequential Standard Model (SSM) \zp\cite{Langacker:2008yv} is used as a benchmark model. In approximately~1~\ifb two subsamples triggered on the presence of an energetic electron or a muon are selected. 
Events with two isolated electrons or muons are kept. 
Fig.~\ref{fig_dileptons_Mll} shows the dilepton mass distribution in both decay channels where the SM backgrounds have been normalized to data in the $Z$ mass region. The test for a presence of a resonance is accomplished using a binned likelihood where systematic uncertainties are treated as nuisance parameters. The systematic uncertainties are due to variations of parton density functions (PDFs), QCD and weak K-factors, as well as trigger, reconstruction and identification efficiencies.
Since the SM model prediction is normalized in the region of the $Z$ pole mass only mass-dependent uncertainties are considered. No significant excess is observed. Upper limits at 95\% confidence level (CL) on cross section times branching ratio ($\sigma B$) 
are computed using \\Bayesian methodology using data from the two decay channels simultaneously. Observed lower limit on the mass of SSM \zp is 1.83~TeV. Lower mass limits are also set on the \esix family of \zp models~\cite{London:1986dk}. Limits  are also set on the $\mathrm{spin\!-\!0}$ KK excitation of the graviton. Shown in Tab.~\ref{table:zpe6gstarlimits} are the lower mass limits on the $E_6$-motivated \zp bosons and RS gravitons.

\begin{figure}
\begin{center}
\resizebox{0.75\columnwidth}{!}{
\begin{tabular}{c}
\includegraphics{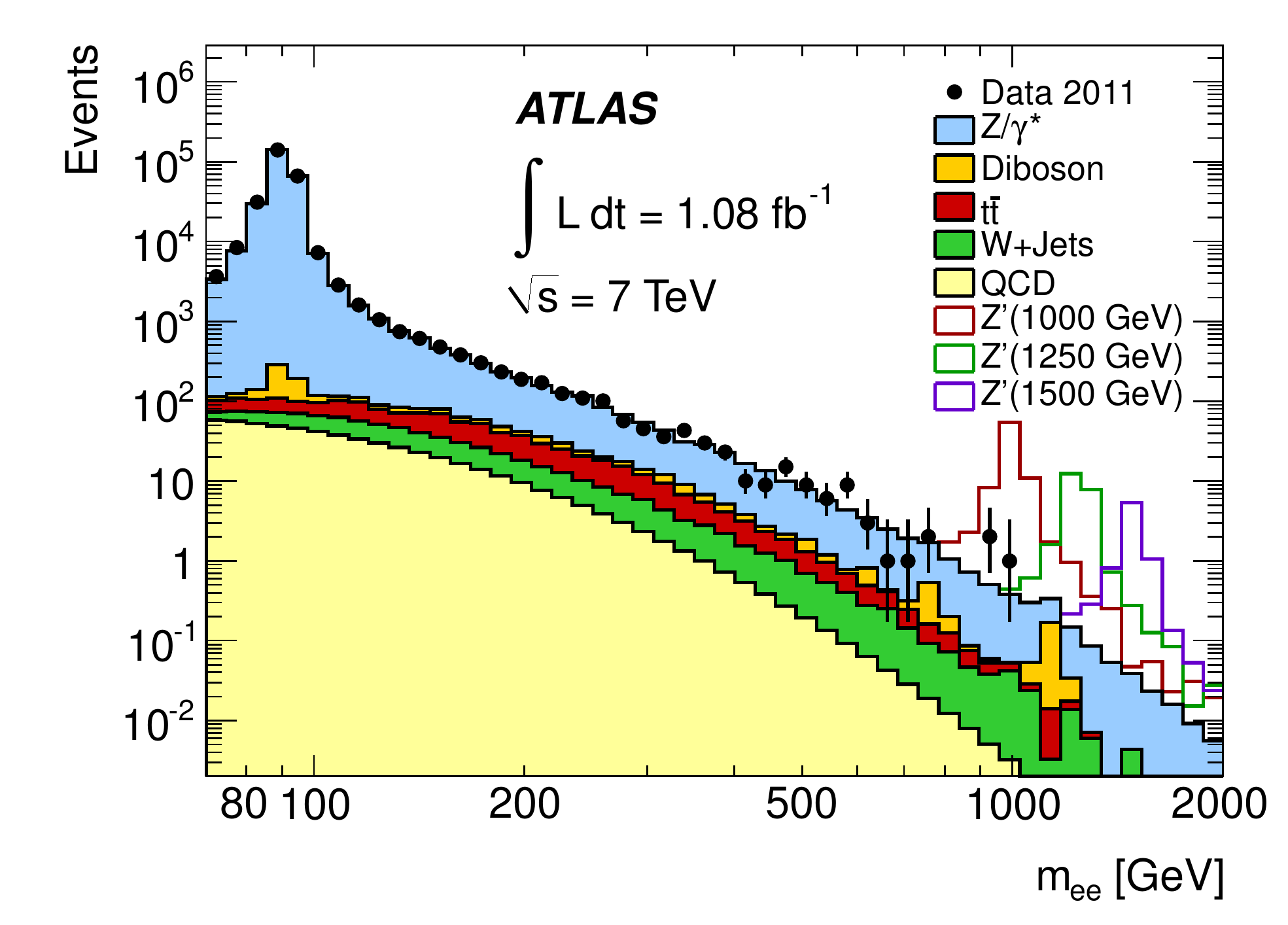}\\
\includegraphics{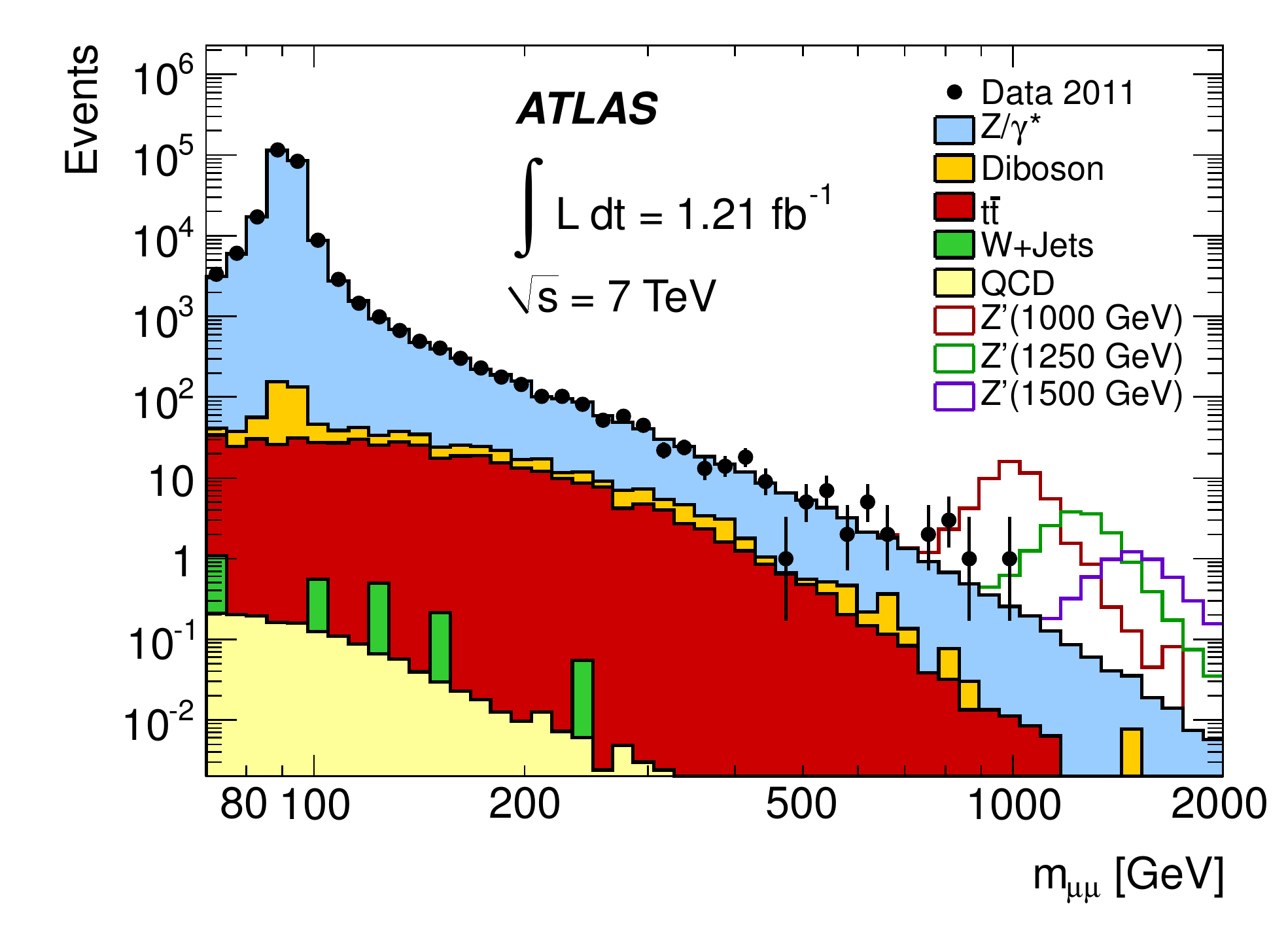}\\ 
\end{tabular}
}
\end{center}
\caption{Dielectron (top) and dimuon (bottom) invariant mass distribution after final selection, compared to the stacked sum of all expected backgrounds, with three example SSM Z' signals overlaid. The bin width is constant in log $m_{ll}$.}
\label{fig_dileptons_Mll}
\end{figure}

%

\begin{table}[!hbt]
\caption{95\% CL lower limits on the masses of \esix-motivated \zp\ bosons and RS gravitons \gstar\ for various values of the coupling $k/\overline{M}_{Pl}$.
Both lepton channels are combined.
}
\label{table:zpe6gstarlimits}
\begin{center}
\resizebox{\columnwidth}{!}{
\small
\begin{tabular}{l|cccccc|cccc}
\hline
& \multicolumn{6}{c|}{\esix\ \zp\ Models}   & \multicolumn{4}{c}{RS Graviton} \\
\hline
Model/Coupling            & \zppsi & \zpN  & \zpeta & \zpI  & \zpsq & \zpchi & 0.01 & 0.03  & 0.05 & 0.1 \\
\hline			 	 	 	 	  	 
Mass limit [TeV] & \LimitCombinedPsi  & \LimitCombinedN  & \LimitCombinedEta  & \LimitCombinedI & \LimitCombinedSq & \LimitCombinedChi & \LimitCombinedGOne & \LimitCombinedGThree  & \LimitCombinedGFive  & \LimitCombinedG  \\
\hline
\end{tabular}
}
\end{center}
\end{table}

The upper limit on the SSM \zp \xbr  is re-interpreted~\cite{US:techihadron} in the context of a techni-meson search. An area in the $m(\rho_T/\omega_T)-m(\pi_T)$ plane excluded by the ATLAS data is shown in Fig.~\ref{fig_dileptons_technihadron_limit}. It is worth pointing out that the excess observed in the CDF data~\cite{cdf:excess} interpreted in the context of technicolor~\cite{Lane:cdfexcess} is ruled out. For a splitting in techni-meson masses of 100 GeV the lower limit on $m_\rho$ is found to be 470 GeV. 
\looseness=-1

\begin{figure}
\begin{center}
\resizebox{0.75\columnwidth}{!}{
\begin{tabular}{c}
\includegraphics{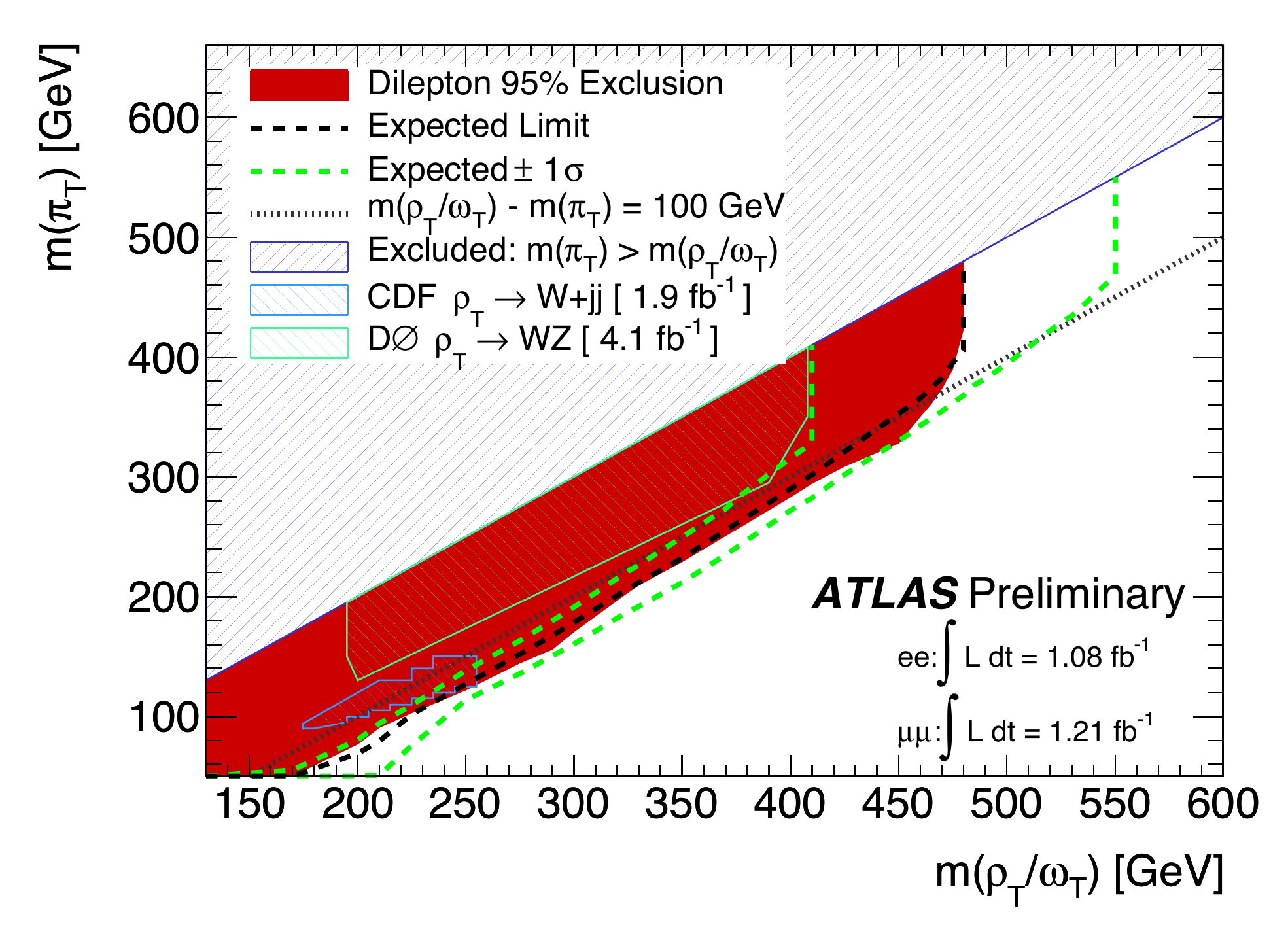}\\
\end{tabular}
}
\end{center}
\caption{The 95\% CL excluded region as a function of the assumed $\pi_T$ and $\rho_T/\omega_T$ masses is shown in red. The dotted line corresponds to $m(\rho_T/\omega_T)-m(\pi_T) = 100 GeV$. The dashed line shows the expected limit with the green dashed lines showing the $\pm 1 \sigma$ bands. The hashed region where $m(\pi_T) > m(\rho_T/\omega_T)$ is excluded by theory. Also shown are recent results from CDF~\cite{CDFtechnisearch} and D\O~\cite{D0technisearch}.}
\label{fig_dileptons_technihadron_limit}
\end{figure}

\section{Search for a heavy diphoton resonance}

The RS graviton branching fraction into a diphoton final state is equal to the combined branching fractions of dimuon and dielectron final states. The search for $\mathrm{spin\!-\!2}$ RS graviton with 2~\ifb is presented in~\cite{US:diphoton}. Data sample was collected using a two-photon trigger. The final sample selection requires presence of two isolated photons
where the invariant mass of the diphoton system exceeds 140 GeV. The invariant mass distribution is shown in Fig.~\ref{fig_diphoton_mgg}. The dominant systematic uncertainties result from PDF variations, data-based procedure of reducible background estimation, and efficiency of photon selection criteria and efficiency extrapolation to high energies. The test for a resonance signal was performed using the {\sc BumpHunter} algorithm~\cite{BumpHunter} and revealed no significant excess. Methodology similar to the one in the dilepton analysis is used to calculate upper 95\% CL limit on the $\gstar \; \sigma B$. 
For $k/\overline{M}_{Pl}$=0.1 $\gstar$ with mass below 1.85~TeV is excluded. Using data from dielectron, dimuon and diphoton final states simultaneously an upper limit is set on the production cross section times the combined branching ratio to the $ee$, $\mu\mu$ and $\gamma\gamma$ final states. The excluded region in the $k/\overline{M}_{Pl}-m_{\gstar}$ plane is shown in Fig.~\ref{limit_grav_diphoton_dil}. For $k/\overline{M}_{Pl}$=0.1 $\gstar\;$ with mass below 1.95~TeV is excluded.
\begin{figure}
\begin{center}
\resizebox{0.75\columnwidth}{!}{
\begin{tabular}{c}
\includegraphics{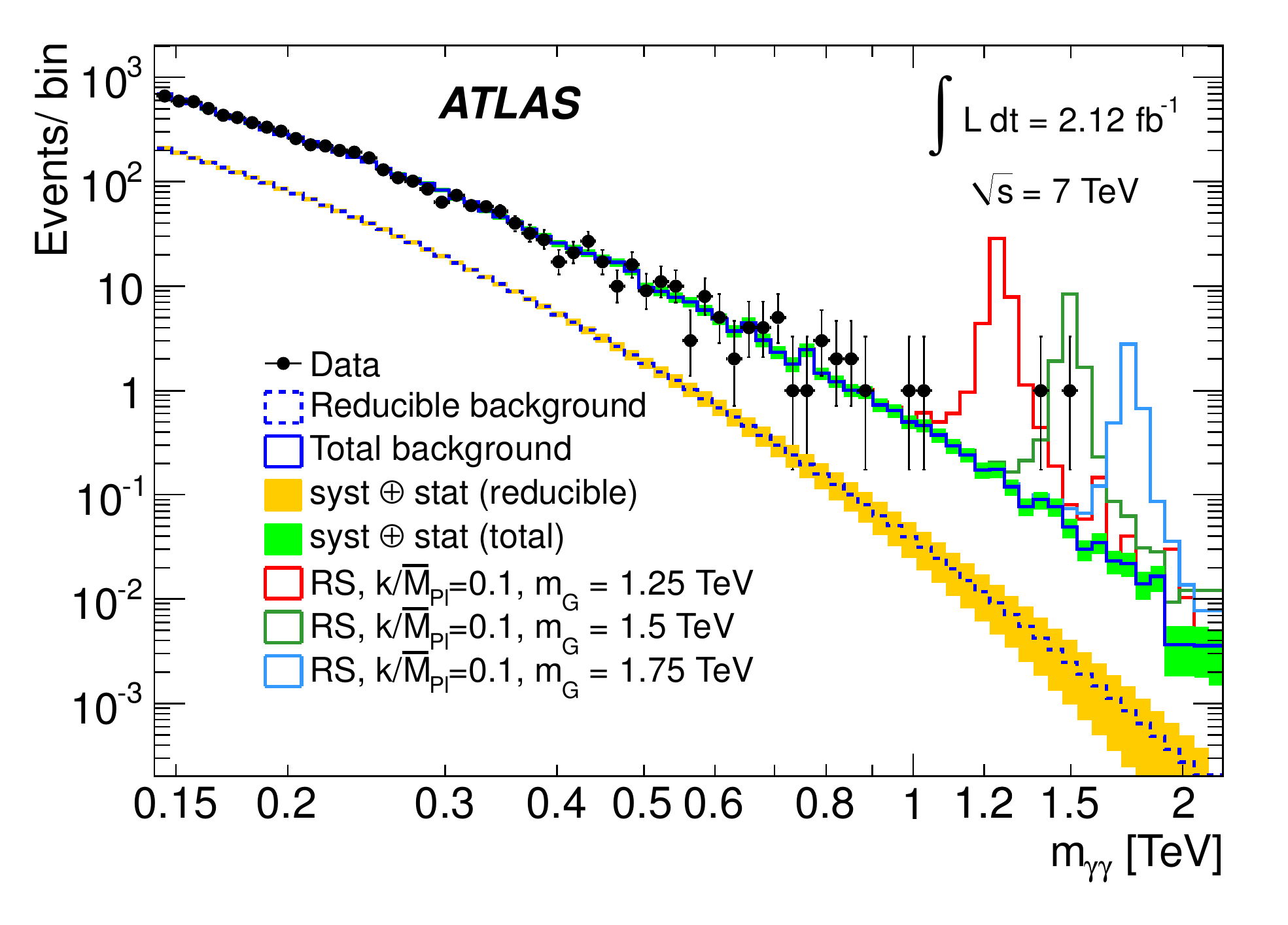}\\
\end{tabular}
}
\end{center}
\caption{The observed invariant mass distribution of diphoton events, superimposed with the predicted SM background and expected signals for RS models with certain choices of parameters. The bin width is constant in $\log(m_{\gamma\gamma})$. The bin-by-bin significance of the difference between data and background is shown in the lower panel.}
\label{fig_diphoton_mgg}
\end{figure}

\begin{figure}
\begin{center}
\resizebox{0.75\columnwidth}{!}{
\begin{tabular}{c}
\includegraphics{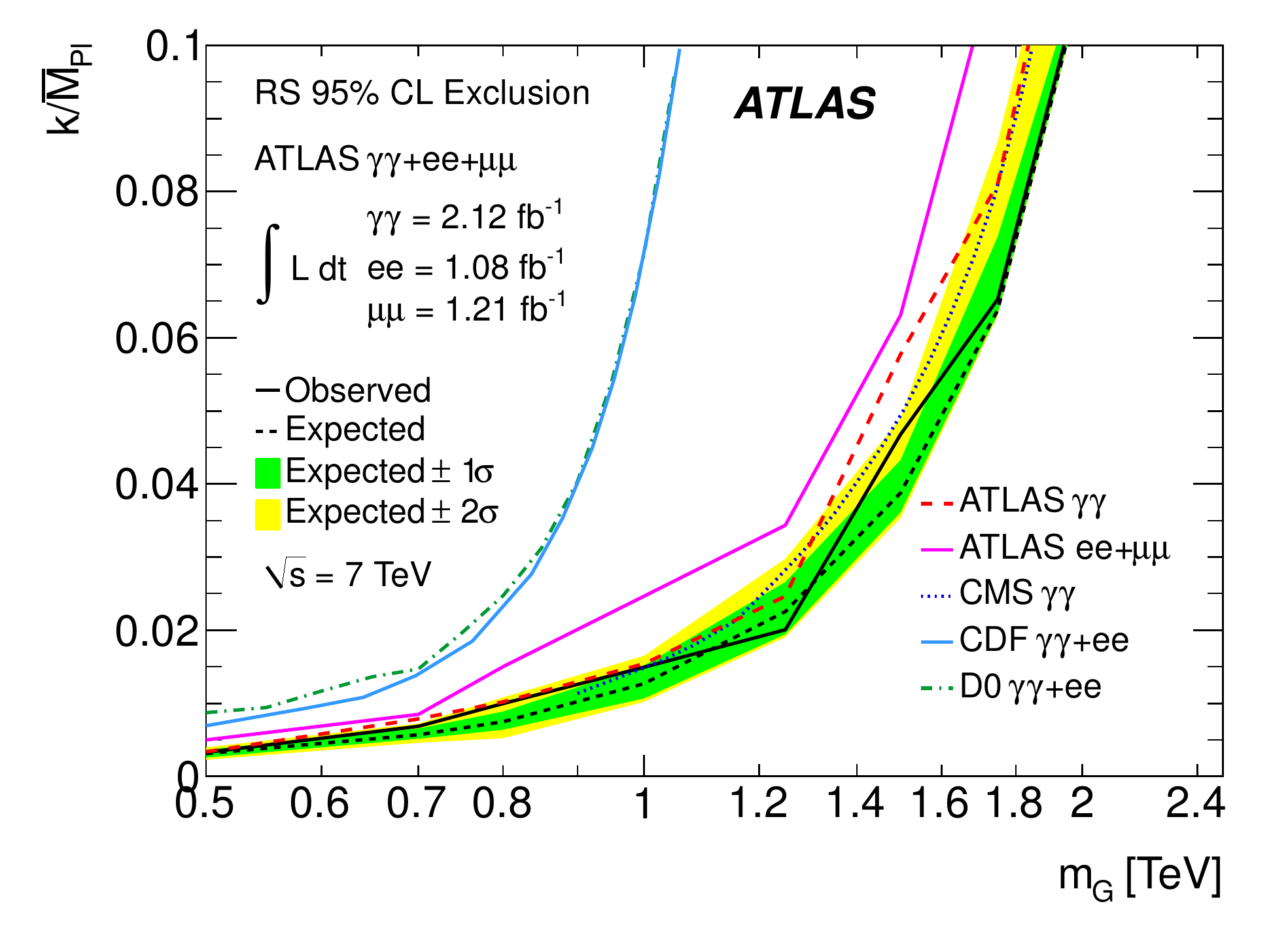}\\
\end{tabular}
}
\end{center}
\caption{The RS results interpreted in the plane of coupling versus graviton mass, and including recent results from other experiments (see references within~\cite{US:diphoton}). The region above the curve is excluded at 95\% CL. In both figures, linear interpolations are performed between the discrete set of mass points for which the dilepton limits were calculated.}
\label{limit_grav_diphoton_dil}
\end{figure}

\section{Search for a heavy gauge boson decaying into a lepton and a neutrino}

In this study~\cite{US:Wprime} two samples of events containing large missing transverse energy $E_T^{miss}$ and exactly one isolated electron or muon are scrutinized for presence of a SSM $W'$ boson~\cite{WprimeSSM}. 
Figure~\ref{fig_WPrime_mT} shows distributions of the transverse mass $m_T$ in electron and muon final states where $m_T=\sqrt{2p_T E_{T}^{miss}(1-cos\phi_{l\nu})}$ and $\phi_{l\nu}$ is the azimuthal angle difference between lepton momentum and the $E_T^{miss}$ vectors. Depending on the $W'$ mass probed a mass threshold $m_{T_{min}}$ is chosen. The discovery significance and upper limit on \xbr is computed using a single bin counting experiment with events with $m_T>m_{T_{min}}$. Bayesian methodology is employed treating the systematic uncertainties as nuisance parameters of the likelihood. The dominant systematics arise from background cross section calculation (including PDF variation), data-driven QCD estimation, momentum scale and resolution modeling, selection efficiency as well as limited size of simulated samples. The SSM $W'$ is excluded at the 95\% level for masses below 2.15 TeV. 
\looseness=-1

\begin{figure}
\begin{center}
\resizebox{0.75\columnwidth}{!}{
\begin{tabular}{c}
\includegraphics{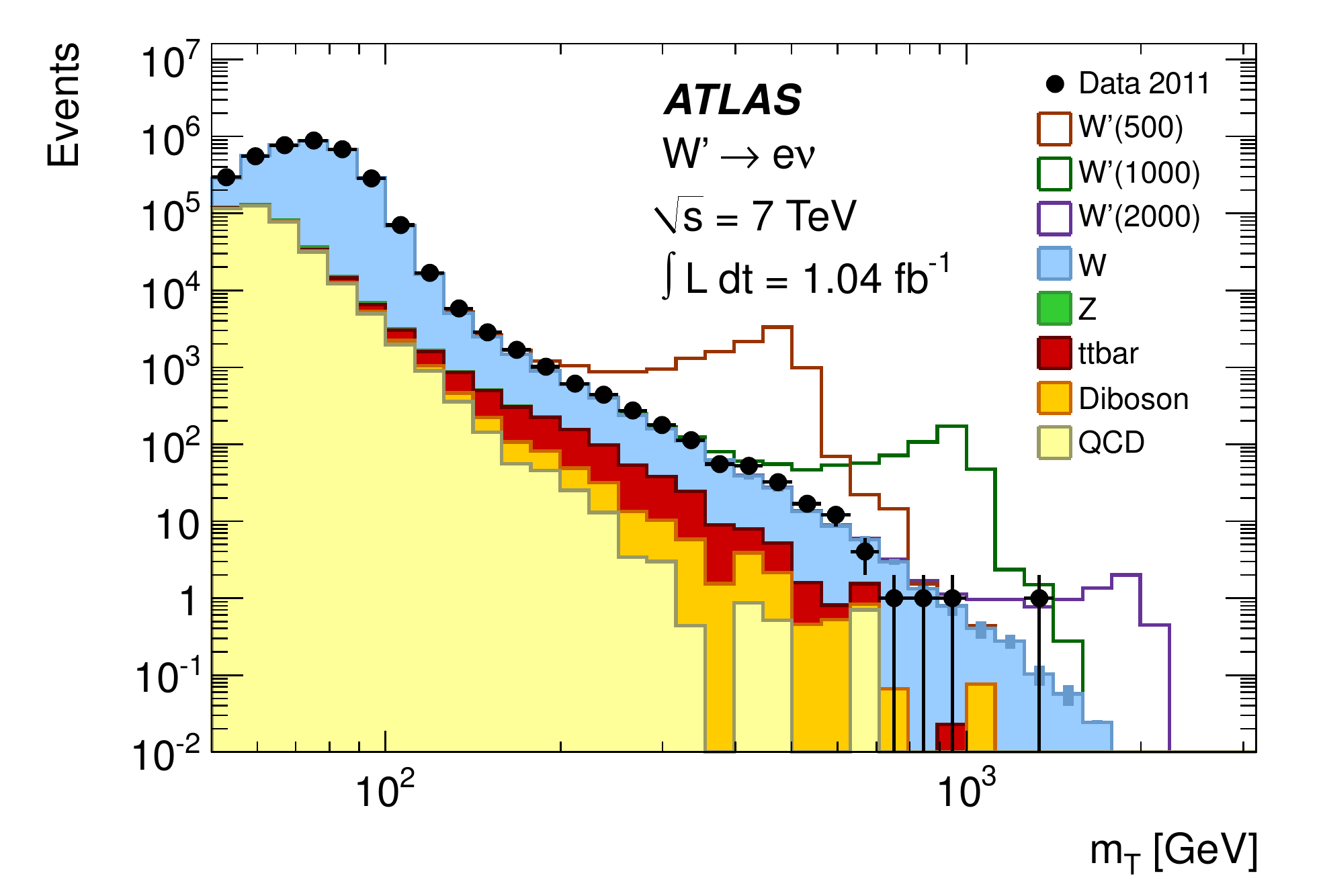}\\
\includegraphics{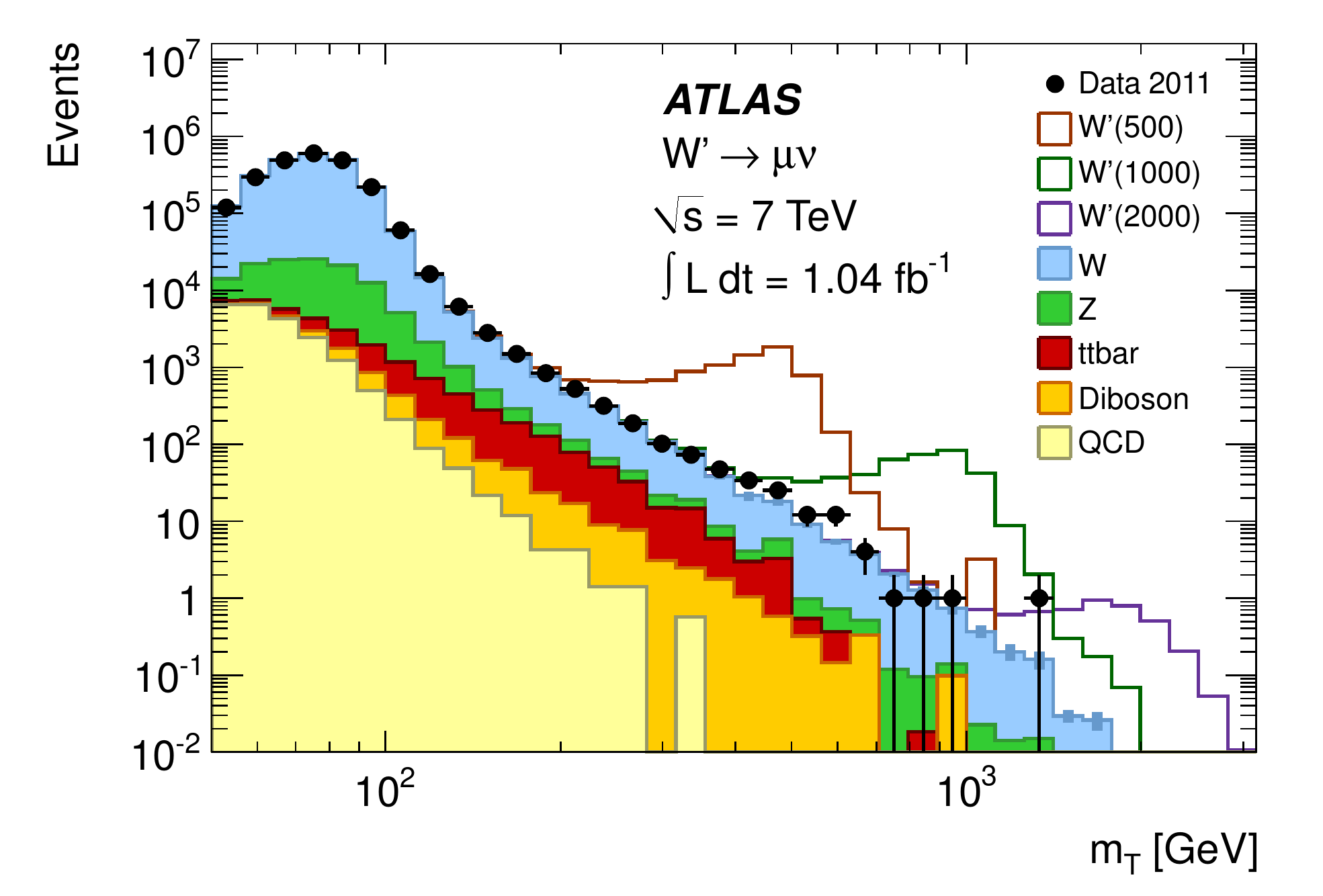}\\ 
\end{tabular}
}
\end{center}
\caption{Spectrum of $m_T$ for the electron channel (top) and muon channel (bottom) after event selection. The points represent data and the filled histograms show the stacked backgrounds. Open histograms are  $W' \rightarrow l\nu$ signals added to the background with masses in GeV indicated in parentheses in the legend. 
}
\label{fig_WPrime_mT}       

\end{figure}


%
\section{Search for a heavy resonance decaying into two jets}

In this analysis~\cite{US:Dijets}, a 1~\ifb sample triggered on a presence of a highly energetic jet is examined for a presence of a dijet resonance. Such resonances are predicted by several extensions to the Standard Model and include the excited quarks $q^*$~\cite{excitedquarks}, axigluons~\cite{axigluons} and color resonances~\cite{s8}. The events with dijet mass $m_{jj}>717$~GeV enter the final analysis sample. 
Kinematic criteria favoring central collisions are applied.

The {\sc BumpHunter} algorithm is applied to search for a resonance-like excess on top of QCD background assumed to behave smoothly. No significant excess is found. In absence of a resonant excess a binned likelihood is used within a Bayesian methodology to set upper \xbr limits on new physics signals. Systematic uncertainties arise due to jet energy scale and luminosity. The excited quarks are excluded below mass of 2.99~TeV, the axigluons below 3.32~TeV and the color octet scalar (s8) below 1.92~TeV. In addition upper limits are set on presence of Gaussian resonance of several relative widths as shown in Fig.~\ref{dijet_limit_gauss}.
\looseness=-1


%
%

\begin{figure}
\begin{center}
\resizebox{0.6\columnwidth}{!}{

\begin{tabular}{c}
\includegraphics{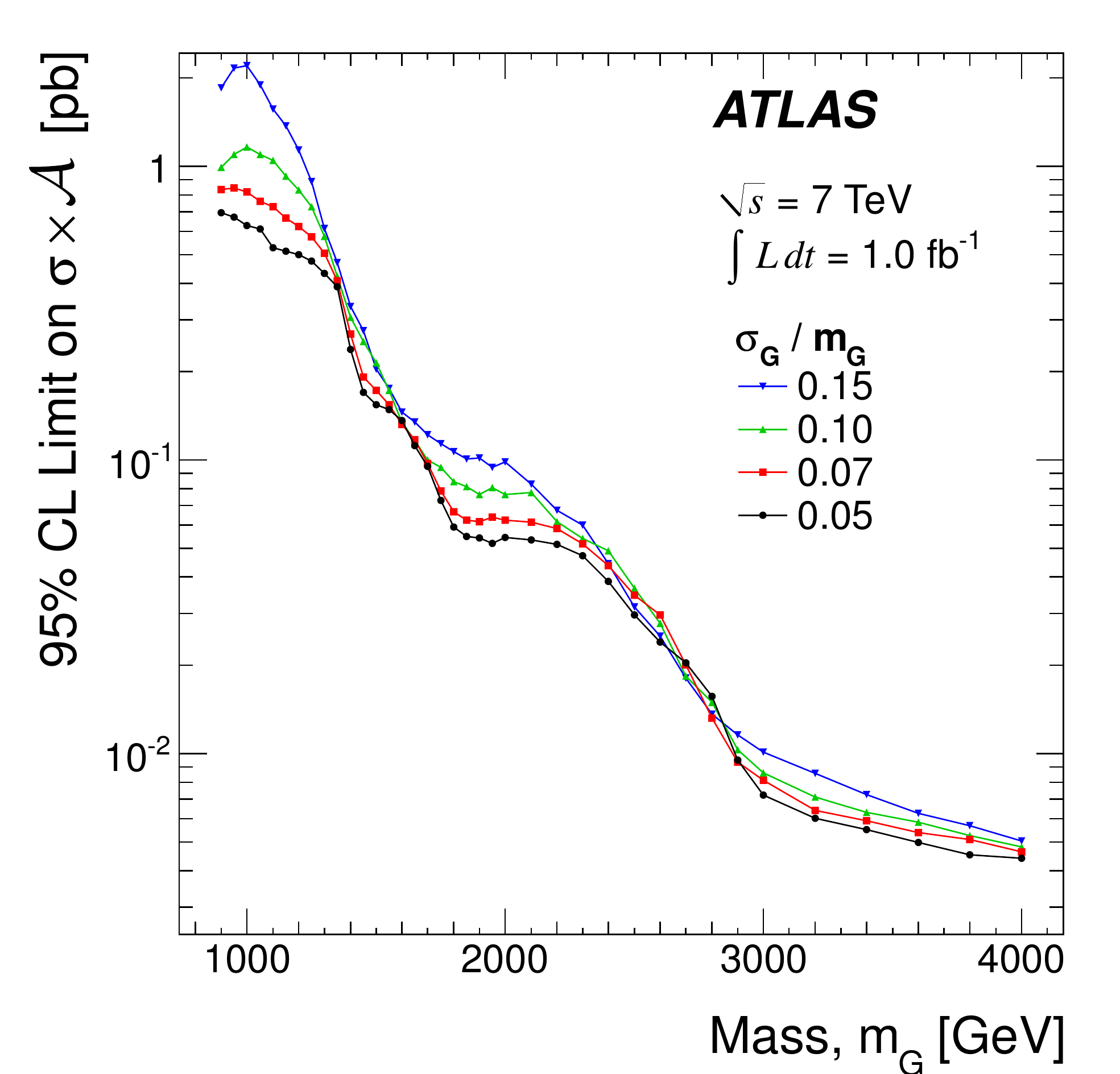}\\
\includegraphics{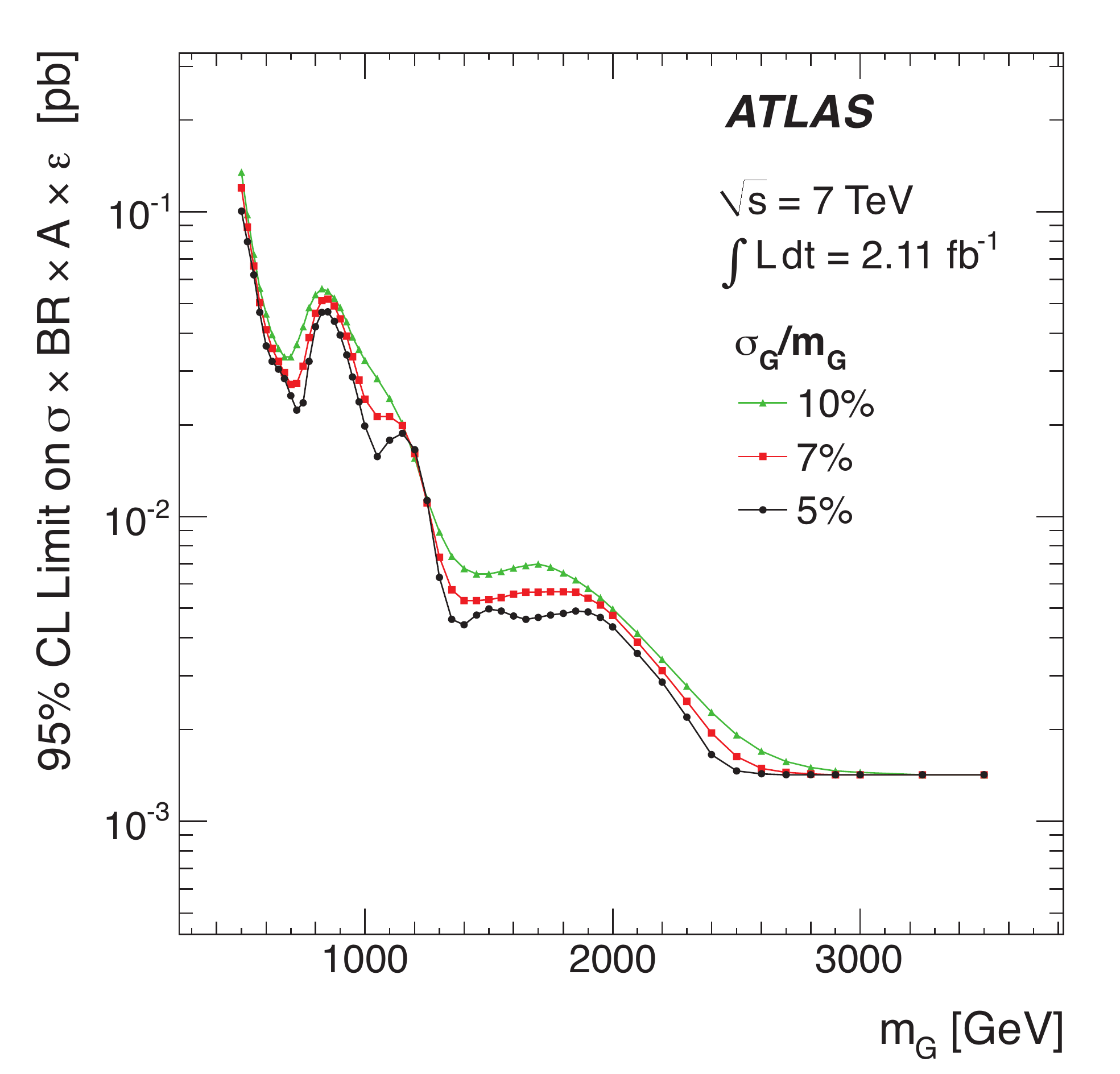}
\end{tabular}

}
\end{center}
\caption{The 95\% CL upper limits on $\sigma \times A$ for a simple Gaussian resonance decaying to dijets (top) and on $\sigma \times B \times A \times \epsilon$ for a Gaussian resonance decaying into a photon and a jet (bottom) as a function of the mean mass, $m_G$, for several values of $\sigma_G/m_G$, taking into account both statistical and systematic uncertainties.}
\label{dijet_limit_gauss}
\end{figure}

\section{Search for Production of Resonant States in the Photon-Jet Mass Distribution}

An analysis employing a similar strategy to the dijet resonance search is the search for resonances decaying to photon-jet final states~\cite{US:photon-jet}. A 2~\ifb sample is analysed. Triggered events require the presence of an energetic photon. In the final analysis the events with at least one isolated photon and at least one jet are considered. The photon must lie in the central region and away from any jets. The distribution of invariant mass $m_{\gamma j}$ of the photon and jet 
is examined using a procedure similar to the one used in the dijet resonance search. No significant excess is found and 95\% CL upper limits 
$q^*$ \xbr. 
Excited quarks with mass less than 2.46~TeV are excluded. Upper limits are also set on production of a generic Gaussian photon-jet resonance as shown in Fig.~\ref{dijet_limit_gauss}
\looseness=-1


%
\section{Searches for $t \bar{t}$ resonances} 

The RS KK gluon $g_{KK}$ excitations~\cite{KKgluon} and other massive particles present in new physics models decay primarily into $\ttbar$ pairs motivating the searches described in detail in~\cite{US:ttbardilepton} in the dilepton channel (1~\ifb) and in~\cite{US:ttbarljets} in the lepton+jets channel (200~\ipb). The $\ttbar$ decay topologies are selected utilizing isolated leptons, jets and missing transverse energy criteria. In the dilepton channel no explicit requirement is made on jet $b$-tagging is made while in the lepton+jets channel one $b$-tagged jet is required. In the dilepton channel the $Z$ boson contribution and non-$\ttbar$ contributions are rejected using dilepton mass cuts and $H_T$ (scalar sum of transverse energies of leptons and jets) cut. In the lepton+jets channel and iterative $dRmin$ algorithm~\cite{US:dRmin} is used to reject the ISR jets contribution. In the lepton+jets channel events with 3 jets are accepted allowing the decay products of the $t$ quarks to be partially merged or fall outside of detector acceptance. 
Systematic uncertainties arise from reconstruction, identification and trigger efficiencies, jet energy scale and resolution, initial and final state radiation (ISR/FSR), parton shower modeling, PDFs, $\ttbar$ event generation using different generators, luminosity and background cross section and $b$-tagging efficiency (in the lepton+jets channel).
The distributions of $H_T+E_{T}^{miss}$ (shown in Fig.~\ref{ttbar_dilep_htMet}) in the dilepton channel and reconstructed $\ttbar$ mass (shown in Fig.~\ref{ttbar_ljets_mass}) are analyzed for presence of a resonant excess. No excess is seen in either of the channels and  $g_{KK}$ \xbr limits are set using Bayesian methodology. in the dilepton channel $g_{KK}$ is excluded up to mass of 0.84~TeV and in lepton+jets channels up to 0.65~TeV. 

\begin{figure}
\begin{center}
\resizebox{0.75\columnwidth}{!}{
\includegraphics{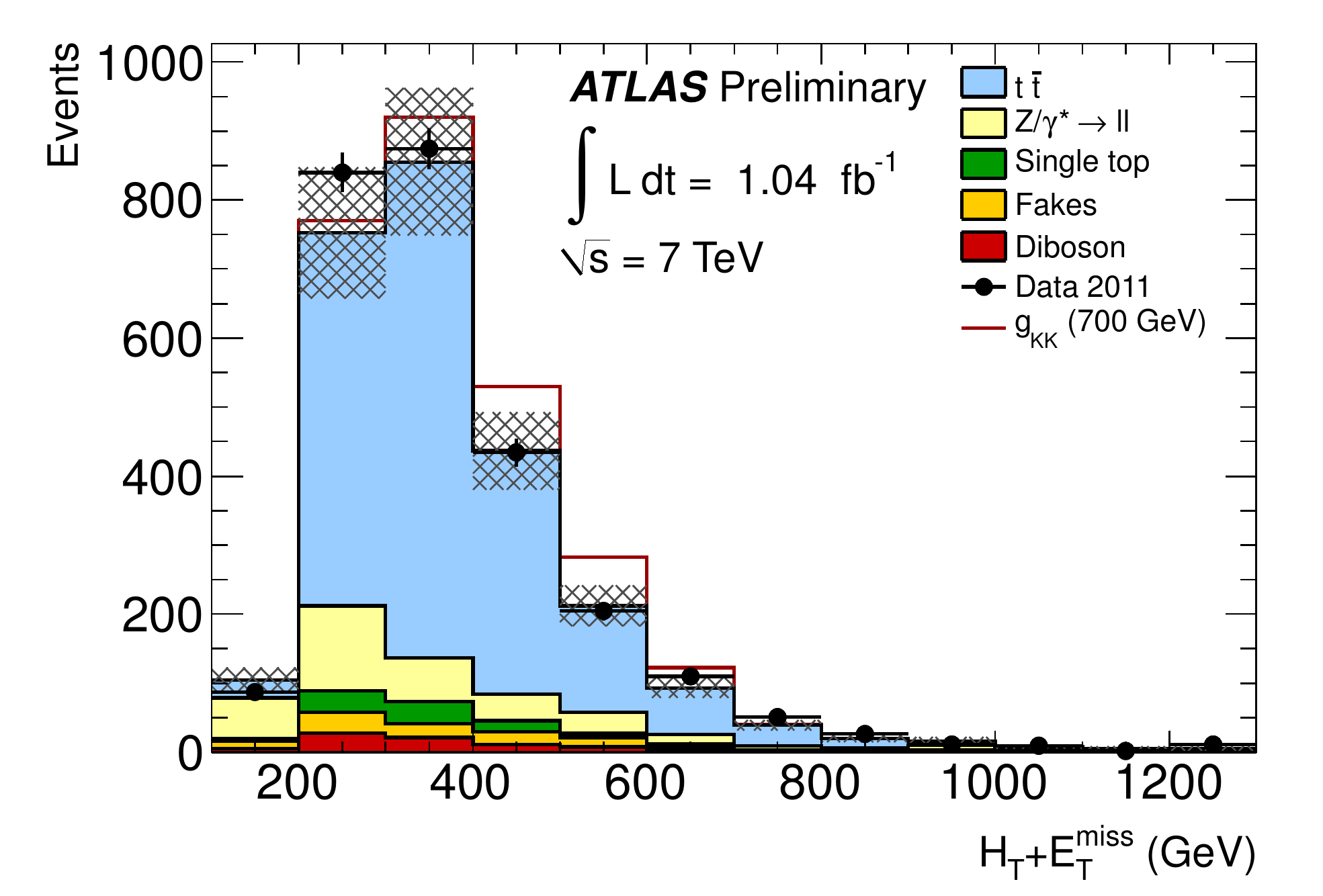}
}
\end{center}
\caption{Data - Monte Carlo comparison for the $H_T + E_{T}^{miss}$ distribution together with a KK-gluon signal with a mass of 700 GeV for illustration. The statistical and systematic uncertainty on the Monte Carlo is represented by the hashed band. }
\label{ttbar_dilep_htMet}
\end{figure}

\begin{figure}
\begin{center}
\resizebox{0.75\columnwidth}{!}{
\includegraphics{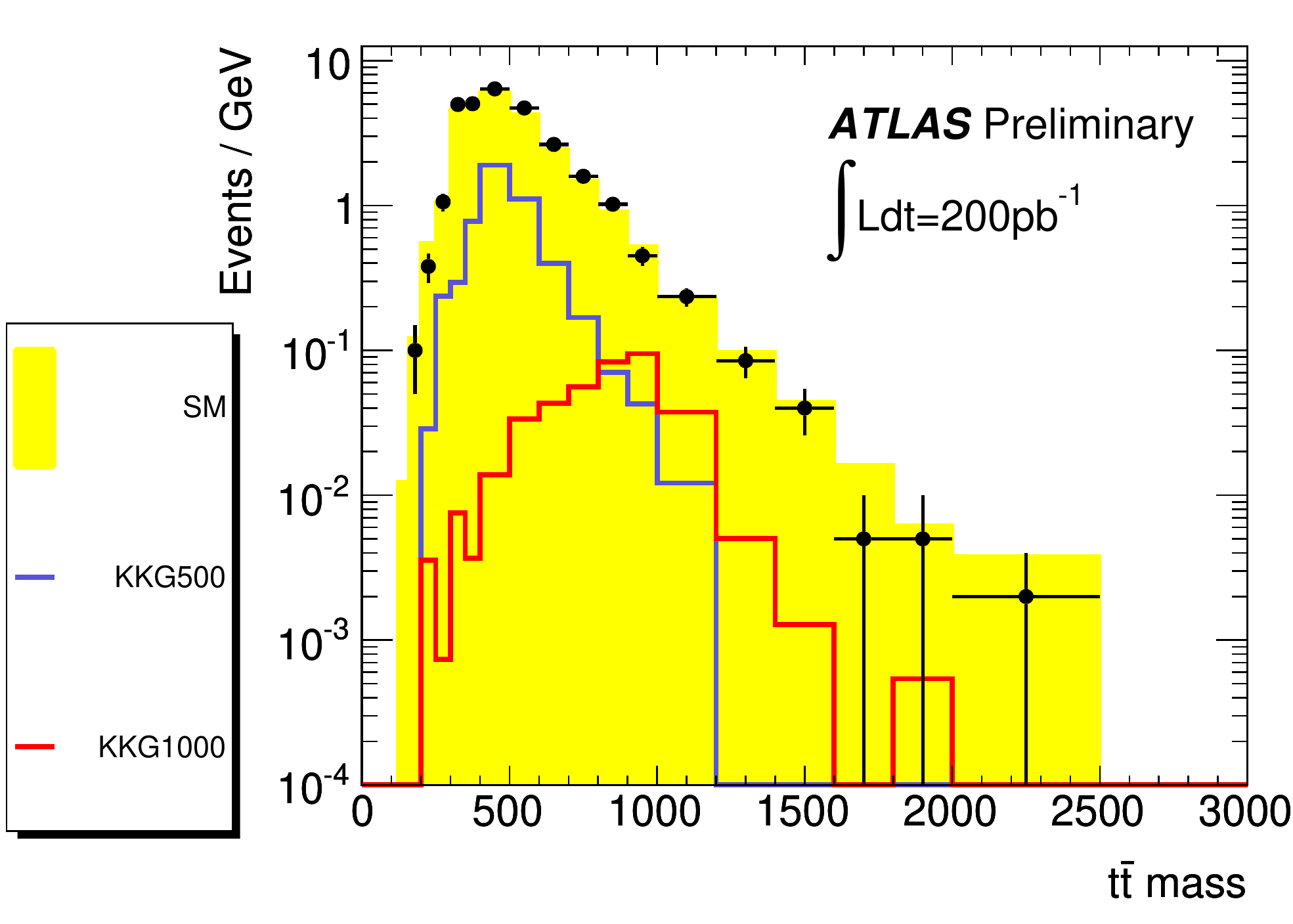}
}
\end{center}
\caption{Reconstructed $\ttbar$ mass  using the $dR_{min}$ algorithm after all cuts. The electron and muon channels have been added together and all events beyond the range of the histogram have been added to the last bin. Only statistical uncertainties are shown.}
\label{ttbar_ljets_mass}
\end{figure}
 
\section{Search for a heavy particle decaying to two $Z$ bosons}

A search for a heavy particle decaying into a pair of $Z$ bosons is presented in~\cite{US:ZZllll}.
Such resonances are predicted for instance in the RS warped extra dimensions model. Events with four isolated leptons are selected. The signal region for this search are the events where leptons form two opposite-sign same flavor pairs. Each pair is required to have the dilepton mass close to the $Z$ boson mass and the four lepton mass $m_{ZZ}$ greater than 300~GeV. The distribution of $m_{ZZ}$ is shown in Fig.~\ref{diboson_llll_m_llll}. The dominant systematic uncertainties arise from the lepton identification and reconstruction uncertainties. No excess in the signal region is observed and the \xbr limits are set using the CLs methodology excluding the RS KK graviton up to mass of 575~GeV for $k/\overline{M}_{Pl}$=0.1. It is worth pointing out that the excess observed at 327~GeV by the CDF experiment~\cite{CDF:ZZllllexcess} is not confirmed.

\begin{figure}
\resizebox{0.75\columnwidth}{!}{
\includegraphics{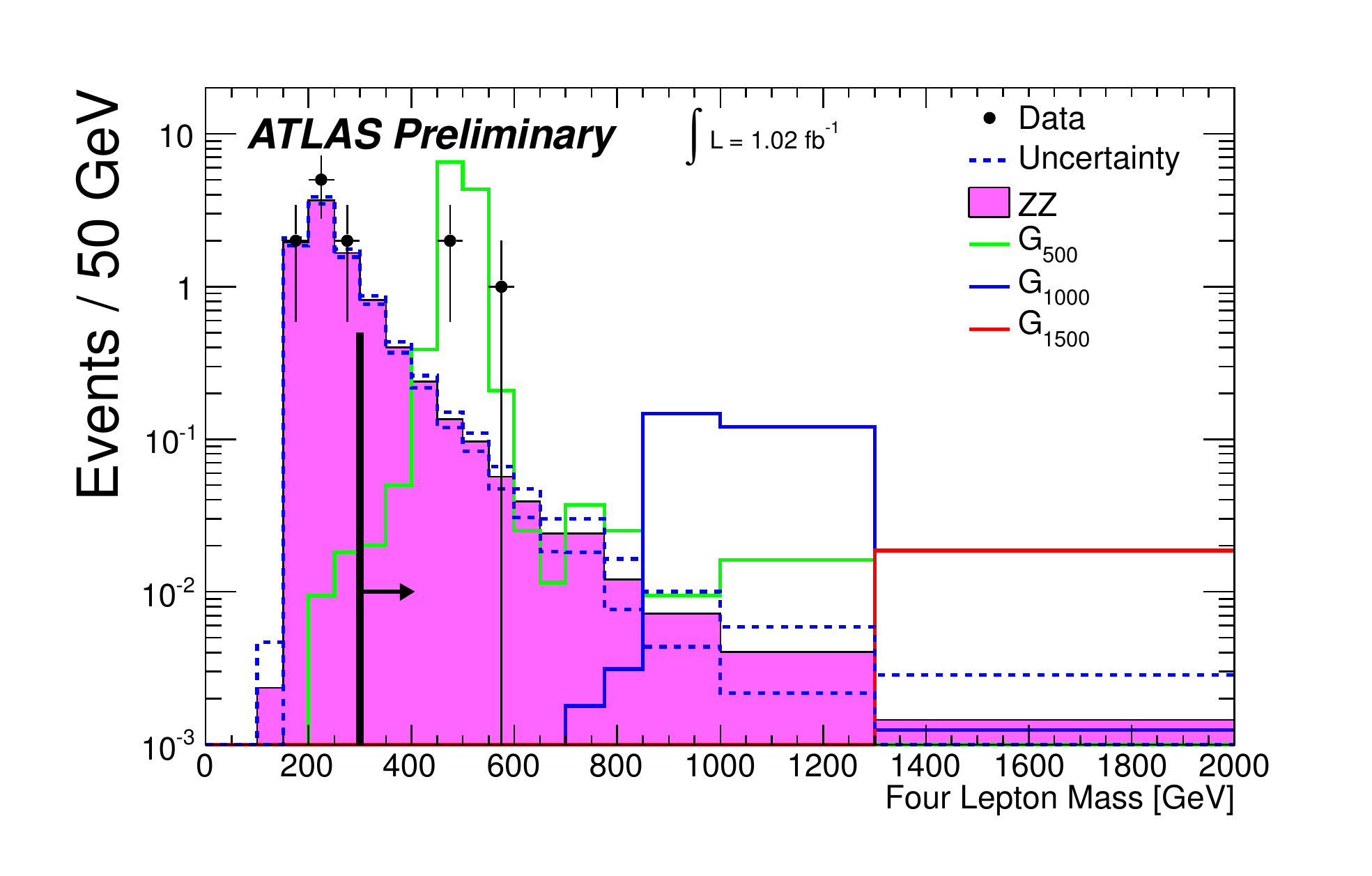}
}
\caption{Distribution of four lepton invariant mass. Fake lepton background is not shown. Hypothetical graviton signal distribution are overlaid. The region with $m_{llll}<$300 GeV, to the left of the solid black line, serves as a $ZZ$ control region; the signal region, indicated by the arrow, is $m_{llll}>$300 GeV. Overflow events are shown in the highest mass bin.}
\label{diboson_llll_m_llll}
\end{figure}

\bibliographystyle{atlasnote}
\bibliography{HighMass}{}


\end{document}
